\documentstyle[psfig,aasms4]{article}
\begin{document}
\title{A search for close bright companions to AeBe stars}
\author{N. Pirzkal}
\affil{Physics and Astronomy Department, University of Wyoming, Laramie WY 82071}
\author{E. J. Spillar}
\affil{US Air Force Phillips Laboratory, Starfire Optical Range, 3550 Aberdeen Av. SE, Kirtland AFB, NM 87117-5776}
\affil{and Physics and Astronomy Department, University of Wyoming, Laramie WY 82071}
\author{H. M. Dyck}
\affil{Physics and Astronomy Department, University of Wyoming, Laramie WY 82071}

\authoremail{norbert@uwyo.edu}
\begin{abstract}
We present the result of IR observations in the K band of 39 bright AeBe stars in the northern hemisphere.  Using the shift-and-add technique, we were able to detect nearby objects as close as 0.4'' from the primary star.  We found 9 stars that have one or more companions.  Adjusting for completeness, we compute that the AeBe star binary rate is 85\%.  The binary rate of AeBe stars is therefore greater than that of near solar type MS stars (57\%) and similar to that found in T Tauri stars (80\%).
\end{abstract}
\keywords{infrared:stars --- stars:pre-main-sequence --- stars:statistics}
\section{INTRODUCTION}
\paragraph{}
AeBe stars are intermediate mass pre-main sequence (PMS) stars ($1.5 < M / M_\odot < 10$) first recognized by \cite{herbig} to be the higher mass equivalents to the T Tauri stars (TTS).  \cite{herbig} defined AeBe stars to be stars of spectral type A or earlier, located in an obscured region, and illuminating a bright nebulosity.  
The number of stars in Herbig's original list was 26, a number later expanded to 57 stars by  \cite{fink}.  While originally broad, the definition of an AeBe star has been steadily refined over the years (see \cite{the} for a good discussion) as the number of AeBe candidates grew. To date, the most comprehensive list of AeBe stars can be found in the work of \cite{the} who published a list of more than 287 AeBe candidates.
\paragraph{}
After an extensive survey of 47 stars \cite{hillenbrand} distinguished among three distinct groups of AeBe stars: stars with large infrared excesses (Group I, candidate disk objects), stars with a flat or rising spectra (Group II, envelope/disk objects), and stars with modest infrared emission (Group III, diskless objects).  \cite{hillenbrand} proposed that the infrared excesses seen in Group I AeBe stars were the result of circumstellar disks.  They showed that the nature of the infrared excesses (specifically a spectral energy distribution (SED) of the form $\lambda F_{\lambda} = \lambda^{-{4 \over 3}}$) could be well modeled by a centrally evacuated ($3<{R_{hole} / R_*}<25$) optically thick circumstellar disk.
\paragraph{}
\cite{hartmann} later argued that the high accretion rates required in the Hillenbrand et al. model to account for the observed amount of infrared light were incompatible with the existence of an evacuated region at the center of the disk.  Instead, relatively small IR excesses could be generated by small dust grains heated in a reflection nebula surrounding the central star, while larger IR excesses might be produced by close stellar companions.  The importance of close nearby objects was explored by \cite{kenyon},  who successfully modeled 1-3$\mu$m infrared light excesses as being produced by deeply embedded stars.
\paragraph{}
To help to resolve this issue, \cite{li} attempted to determine the binarity of AeBe stars by obtaining quasi-simultaneous JHK photometry of 19 AeBe stars.  They used small apertures to avoid contamination from nearby sources and found that 9 stars had a companion star within 10 arcseconds of the primary with separations ranging from 2500 to 9000 AU.
Their search was however restricted by atmospheric seeing to objects farther than 1.1'' - 1.6''.  For an AeBe star at a distance of 2500 pc, such as MWC 1080,  sources closer than 2750 AU to 4000 AU would not have been detectable.
\paragraph{}
High resolution observations can be achieved from a ground based telescope using the shift-and-add (SAA) technique (\cite{bates}).  SAA is a simple technique which consists of taking a series of very short exposures (specklegrams) that freeze the image of the star before it is blurred across the detector.  A long exposure image, free of atmospheric tilt errors, can then be constructed by individually shifting the image of the star to the middle of each specklegram before co-adding them together.  
This technique has been shown to work well in the past and to retrieve most of the high spatial frequencies usually lost to seeing (\cite{christou1}).
\paragraph{}  
Our experience with SAA at a 2.3m class telescope has shown that we are typically able to reduce the FWHM of a point source by a factor of 3 in the near infrared.  On observing nights with an average seeing we repeatedly managed to construct high resolution images with a FWHM of 0.3'' - 0.4''.
\paragraph{}
We conservatively estimate our inner detection limit at 0.4'' and maximum K magnitude for an observable secondary at K = 10.5 (with S/N of 10 in the final shift-and-added frame).  This is to be compared to the angular resolution limit by Rayleigh's criterion of 0.24''. 

\section{OBSERVATION}
\paragraph{}
The data presented here were obtained in July and December 1995 and July 1996 using the WIRO 2.3m telescope and a 256x256 NICMOS3 infrared camera equipped with the approximate Johnson J, H, and K filters.  The image scale was .12''/pixel and we only used one quadrant of the detector at a time resulting in a field of view of 15''.  In July and December 1995, each observation consisted of a series  of one hundred 60 ms specklegrams obtained in less than a minute, thus ensuring that the point spread function (PSF) of the system did not change during the course of the observing.  The data were taken in the same manner in July 1996 but using larger series of specklegrams, each containing 200 frames.
\paragraph{} 
Due to our short integration times, the readout noise of the camera was the dominant source of noise.  Hence, since we required a S/N ratio high enough to accurately determine the position of the primary star in each specklegram, we were only able to observe objects brighter than $9^{th}$ magnitude (in the K band).  The S/N of secondary objects in the field of view did not have to be as high as demonstrated by the fact that we were able to detect companions as faint as 10.5 magnitude.
\paragraph{} 
We observed 39 of the brightest AeBe stars listed by \cite{the} over a period of one year in the K band (Table 1).  Each program star was preceded and immediately followed by a point source star to serve as an estimate of the PSF.  This was done twice using different quadrants of the detector.  The specklegram series was then shift-and-added to produce high resolution images of the program stars and point source objects.

\section{DISCUSSION}
\paragraph{} 
We were able to identify 9 multiple systems (7 binaries and 2 triple systems) with angular separations ranging from .6'' to 7'' (figure 1).  For statistical reasons we differentiate between two types of stars: those with a reasonable distance estimate (27 stars), and those with no distance estimate (12 stars).  Among the first group we found 7 multiple systems with projected physical separations ranging from 176 to 11725 AU.  Two additional multiple systems were found in the second group (HD 216629 and HD 141569).
Previously observed companions around HD 200775 (\cite{li}), and around XY Per (\cite{HBC}) were successfully detected.  The separation and position angle of  the detected multiple systems are given in table 2. The close companions around MWC 1080, XY Per, and HD 150193 were successfully resolved using the CLEAN algorithm (\cite{hogbon}), using the image of a nearby star to serve as an estimate of the PSF.   The two companions surrounding XY Per are an artifact from the shift-and-add caused by the fact that XY Per and its companion are close and are of nearly equal brightness.
\paragraph{}
Using these observations, it is possible to determine the mean number of companions for AeBe stars and to compare it to that of TTS (\cite{ghez}) and near solar type MS stars (\cite{duq}).
For each object in our list, a Monte Carlo technique was used to generate and project onto the background sky ten thousand binary systems with all possible orientations (i, $\omega$, $\Omega$) and configurations (a, e, q=$M_2$/$M_1$).  These were used to determine the number of times a companion would be visible around a given AeBe star.  In order to be visible, a nearby companion had to be between 0.4'' and 8'' (conservative estimates of the resolution limit and of half of the field of view, respectively) from the primary star and had to be massive enough to be visible in the K band (we assumed main sequence color corrections, \cite{allen}).  The eccentricity and mass ratio distributions were taken from \cite{duq}.
The masses of the AeBe primaries were not known and we averaged the results obtained using a flat distribution of masses ranging from $1.5 M_\odot$ to $10 M_\odot$.  When the distance to an AeBe star was unknown, we additionally averaged the results obtained by placing the star at distances ranging from 100pc and 1500pc.  Table 1 shows the thus predicted binary discovery rate of each of the stars in our sample. 
\paragraph{}
By assuming an intrinsic binary frequency of 100\%, we computed that 16\% $\pm$1.6\% of the companions would be detected or 6.2 $\pm$0.6 stars for a sample of 39 binary stars.  We also calculated that 0.2 field stars brighter than 10.5 K magnitude would be observed within 8'' (based on computations using the \cite{soniera} model). Adjusting for fortuitous field stars, we therefore expected to detect 6.4 $\pm$0.6 multiple systems if all of the 39 stars in our sample were binaries.  The successful detection of 9 multiple systems out of 39 AeBe stars corresponds to a mean number of companions of 1.4.  Taking into account the uncertainty in our monte-carlo prediction and Poisson statistics we calculated that there is a 95\% likelihood that the AeBe star binarity is greater than 85\%.  This is comparable to that of TTS ( 80\% $\pm$22\%  \cite{leinert}, \cite{ghez}) and exceeds that of near solar type MS stars (57\%,  \cite{duq}).  Note that sources were previously detected around some of these AeBe stars (Z CMa (\cite{haas}), VV Ser and V376 Cas (\cite{li}) for example), but that these were not combined with our observations because they were not within the completeness region of this study (0.4''$<$sep.$<$8'' and $K<$10.5).  It should be noted that to obtain a significant increase in the confidence level would require a sample at least four times as large.  Such a large sample is not available in the northern hemisphere, but this statistical study can be improved by observing additional AeBe stars in the southern sky.
\paragraph{}  
It is also important to note that we found companions (which are assumed to be the fainter stars) around 30\% of the type I AeBe stars  in our sample (HD 200775, BD+46 3471, MWC 1080, KK Oph, and HD 150193).  In the case of HD 200775 and BD+46 3471 the companions were found to be significantly fainter ( $\Delta$K $=$ 4.9 and 4.7 respectively) than the primary star and can only account for a small portion of the observed infrared light excess (\cite{hillenbrand}).  The nearer of the two companions around MWC 1080 contributes about 10\% of the 2.2 $\mu$m radiation and we estimate that the companions around KK Oph and HD 150193 contribute about 10\% and 13\% of the total 2.2 $\mu$m radiation.   The companion near V892 Tau, the only type II star with a companion in this study, is seen to contribute less than 1\% of the total 2.2 $\mu$m light.  The three stars that have not previously been classified by \cite{hillenbrand} have been found to all have very bright companions contributing between 20\% and 100\% of the total infrared light from the system.
\paragraph{}
The ambiguity as to which star is the primary star and which one is the secondary cannot be resolved without further observations of these systems using different wavelengths.  However, we can measure the K magnitude differences between the various components of the observed multiple systems and compare those to the 2.2 $\mu$m infrared excess estimates published by \cite{hillenbrand}.  The measured $\Delta$K of MWC 1080, V892 Tau, and KK Oph are comparable to the estimates from \cite{hillenbrand}.  The infrared excess in those systems could therefore be caused by the observed stars if the primary AeBe star is actually the dim star in those fields.  The $\Delta$K measured in the remaining three systems for which an estimate of the infrared excess is available are an order of magnitude too small and none of these companions are likely to be the source of the infrared excesses. 
No stars were found around other type I stars even in stars such as MWC 275 and AB Aur which are relatively close by and have been observed to have a large infrared excess (\cite{hillenbrand}).  We believe that the relatively close distance of some of these type I AeBe stars and the fact that no star was found close to them make them good candidates for the search for an extended structure.

\section{CONCLUSION}
We have successfully used the shift-and-add technique to obtain high resolution images of AeBe stars.  We observed 39 AeBe stars and determined the binarity of AeBe stars to be high (85\%) based on the detection of 9 multiple systems.   
While none of the companions seem bright enough to be a major source of infrared excess, we showed that the 2.2 $\mu$m infrared excess of three of the observed systems could be explained if we assume that the dim star is the primary AeBe star. The infrared excess in the remaining systems cannot be attributed to any of the observed companions.

\section{ACKNOWLEDGMENT}
Dr. E. J. Spillar would like to acknowledge the support of the US Air Force during his sabbatical at the Starfire Optical Range.
We would like to thank Jeff Kuhn of Michigan State University and the National
Solar Observatory and Steve Boughn of Haverford College for helping to 
build the camera.

\newpage
\centerline{\psfig{figure=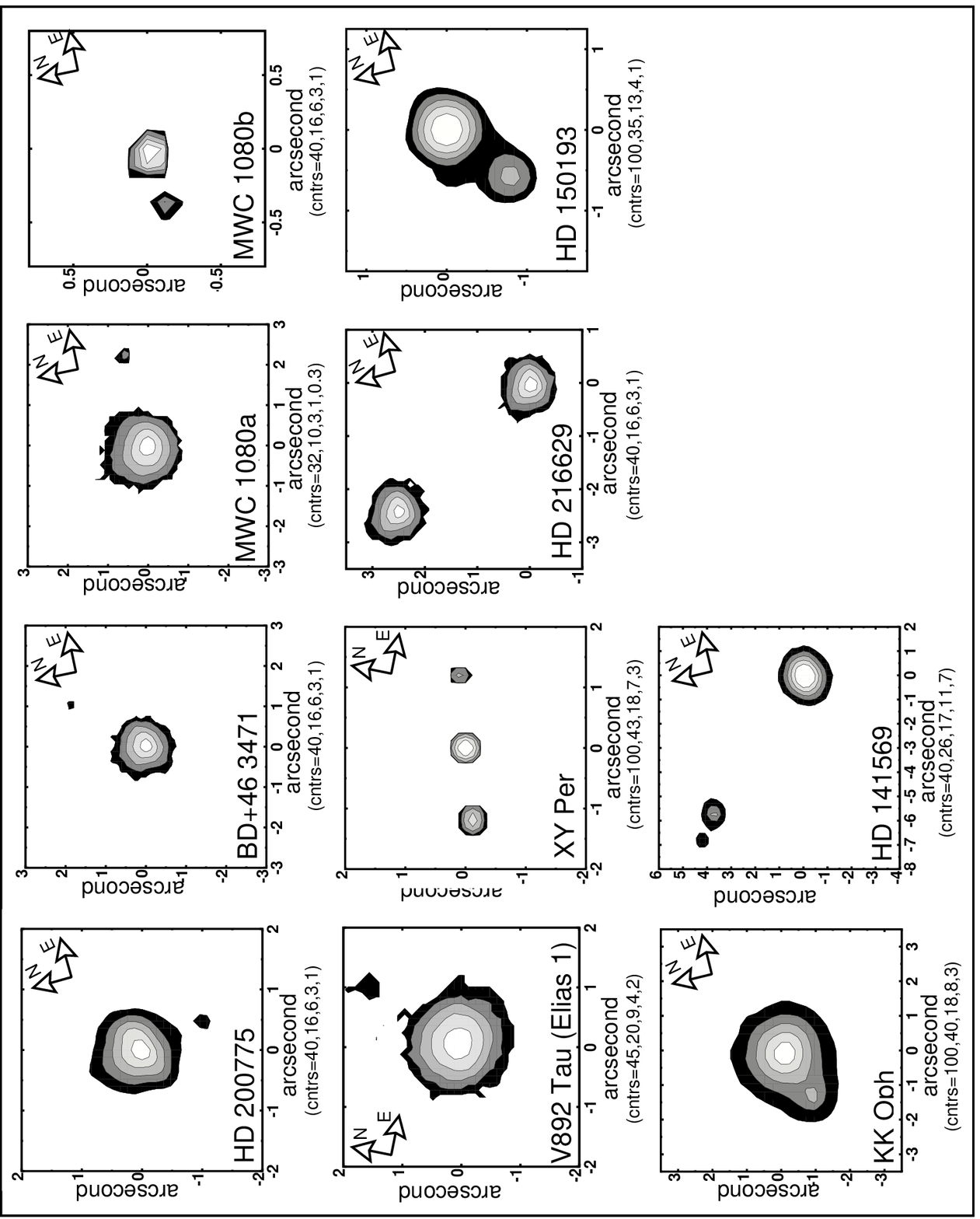,width=5.0in}}
\figcaption[fig1.1.eps]{Contour plots of the nine multiple AeBe systems: HD 200775, BD+46 3471, MWC 1080, V892 Tau, XY Per, HD 216629, HD 150193, KK Oph, and HD 141569.  The contour levels are arbitrary and normalized to 100.  The third component of the XY Per system is an artifact of the shift-and-add technique. \label{fig1}}

\begin{deluxetable}{lcccccc} 
\tablecaption{AeBe Stars}
\tablewidth{0pt}
\tablehead{
\colhead{Star}   &
\colhead{Group\tablenotemark{1}}       & \colhead{D\tablenotemark{2}}   & \colhead{Ref.} 
& \colhead{\sl{q min}\tablenotemark{3}}  
& \colhead{Disc. Rate\tablenotemark{4}}\nl
& & (pc) &D&($M_2$/$M_1$)&(\%)
}
\scriptsize
\startdata
VV Ser & I &440&6 & .2 &18\nl
MWC 614  & \nodata & \nodata & \nodata& .2 &17 \nl
WW Vul & \nodata & 550& 7 & .35 &13  \nl
V1295 Aql & \nodata& \nodata &\nodata&.15 &18 \nl
AS 442 & \nodata& 700&8 &.2 &17 \nl
V1493 Cyg & \nodata&\nodata&\nodata & .35 & 12 \nl
HD 200775 & I&600&6 & .15 &19\nl
V645 Cyg & \nodata&6000 & 12& .2 &8\nl
V373 Cep & I& 1000&5 & .3 &13\nl
BD+46 3471 & I& 900&6  & .2 &16\nl
HD 216629 & \nodata&725&12 & .3 &14 \nl
MWC 1080 & I& 2500&6 & .3 &10\nl
V376 Cas & II& 600&6 & .3 & 14\nl
VX Cas  & \nodata&\nodata &\nodata& .2 &17\nl
V594 Cas & I& 650&9 & .2 &17\nl
IP Per &  \nodata&350&10 & .4 &13\nl
XY Per & \nodata&160&12 & .2 &22\nl
V892 Tau\tablenotemark{5} & II& 160&5 & .2 & 22\nl
AB Aur & I& 160&6 & .1 &26\nl
MWC 480 & \nodata&\nodata &\nodata& .2 &17\nl
UX Ori & I& 460&6 & .3 &15\nl
MWC 758 & \nodata&\nodata &\nodata &.2 &17\nl
CQ Tau & \nodata&\nodata &\nodata &.2 &17\nl
HK Ori & I& 460&6 & .3 &15\nl
HD 244604 & \nodata&\nodata & \nodata&.3 &14\nl
Z CMa & \nodata&1150&9 & .1 &17\nl
MWC 166 & III& 1150&6 & .15 &16 \nl
KK Oph & I & 310 &12 &.2 &17\nl
HD 141569 & \nodata & \nodata & \nodata&.3 &14 \nl
HD 144432 & \nodata & \nodata &\nodata &.37 &12\nl
HD 203024 & \nodata & \nodata &\nodata& .35  &12 \nl
LkHa 134 & \nodata & 700&12 & .3 &14 \nl
MWC 275 & I  & 125&11&.2  &22\nl
MWC 297 & I & 450&12 &.05 &23\nl
MWC 300 & \nodata  & 700&8 &.2  &17\nl
HD 150193 & I & 160 & 6 &.28 &14 \nl
V1685 Cyg & I & 1000&8  &.1 &18\nl
V1686 Cyg  & I & 1000&8  & .15 &16\nl
V361 Cyg  &III & 1000&9  &.2 &15
\enddata
\tablenotetext{1} {As defined by \cite{hillenbrand}}
\tablenotetext{2} {Distance of object from the literature}
\tablenotetext{3} {Minimum mass of a detectable companion.}
\tablenotetext{4} {Probability of detecting an existing secondary around the star.}
\tablenotetext{5} {V892 Tau listed as Elias 1 in \cite{hillenbrand}}
\tablerefs{(6) \cite{hillenbrand}
;(7) \cite{brooke};(8) \cite{terranegra};(9) \cite{fink};(10) \cite{wood};(11) \cite{blondel};(12) \cite{berrilli}
}
\end{deluxetable}

\begin{deluxetable}{lccccc} 
\tablecaption{Detected Multiple AeBe Systems}
\tablewidth{0pt}
\tablehead{
\colhead{Star} & \colhead{Sep.} &\colhead{Pos.Angle} & \colhead{\sl{p.s}\tablenotemark{1}}&$\Delta$K\tablenotemark{2}&$\Delta$K\tablenotemark{3}\nl
& (arcsec) &(degrees) &(A.U.)
}
\scriptsize
\startdata
HD 200775 & 2.25 & 164  & 1350 &4.9 & 1.5\nl
BD+46 3471 &  4.18 & 43  & 3762 &4.7 & 1.0\nl
HD 216629 & 6.96 & 147  & \nodata & 0.0 & \nodata\nl
MWC 1080 & 4.69 & 86  & 11725 &6.2 & 2.76\nl
& .6 & 270 & 1500 &3.1 & 2.76\nl
XY Per & 1.2 & 255   &192&0.0 & \nodata\nl
V892 Tau\tablenotemark{2} & 3.72 & 19  & 595 &4.2 & 5.0\nl
KK Oph & 1.5 &257 & 465 &2.5 & 3.0\nl
HD 141569 & 8&314 &\nodata &2.0 & \nodata\nl
& 6.8 & 312 & \nodata&2.7 & \nodata\nl
HD 150193 & 1.1& 236 &176 & 2.2 & 5.0\nl
\enddata
\tablenotetext{1} {Projected separation at distance \sl{D} given in Table 1}
\tablenotetext{2} {K magnitude difference between primary and companion star}
\tablenotetext{3} {From \cite{hillenbrand}, infrared excess at 2.2 $\mu$m in astronomical magnitude}
\tablenotetext{4} {V892 Tau listed as Elias 1 in \cite{hillenbrand}}
\end{deluxetable}
\end{document}